\documentclass[aps,prb,reprint,showpacs,superscriptaddress]{revtex4-1}

\usepackage{graphicx}
\usepackage{amsmath}
\usepackage{amssymb}
\usepackage{epsfig}

\renewcommand{\v}[1]{{\bf #1}}
\newcommand{\sign}{{\rm sign}}

\newcommand{\psib}{{\bar{\psi}}}

\newcommand{\w}{{\omega}}

\def\eqa{\begin{eqnarray}}
\def\eea{\end{eqnarray}}
\newcommand{\eq}{\begin{equation}}
\newcommand{\ee}{\end{equation}}
\newcommand{\nn}{\nonumber\\}

\newcommand{\<}{\langle}
\renewcommand{\>}{\rangle}

\newcommand{\p}{\partial}

\newcommand{\ua}{\uparrow}
\newcommand{\da}{\downarrow}
\newcommand{\ra}{\rightarrow}

\newcommand{\al}{\alpha}

\newcommand{\del}{\delta}
\newcommand{\Del}{\Delta}
\newcommand{\eps}{\epsilon}

\newcommand{\Ga}{\Gamma}

\newcommand{\la}{\lambda}
\newcommand{\La}{\Lambda}

\renewcommand{\th}{\theta}

\newcommand{\si}{\sigma}

\newcommand{\cP}{ {\cal P} }

\newcommand{\cT}{ {\cal T} }

\usepackage{color}
\begin{document}
	
\title{Spin-triplet $f$-wave pairing in twisted bilayer graphene near 1/4 filling}
	
\author{Qing-Kun Tang}
\affiliation{National Laboratory of Solid State Microstructures \& School of Physics,
	Nanjing University, Nanjing, 210093, China}

%\author{Weicheng Bao}
%\affiliation{National Laboratory of Solid State Microstructures \& School of Physics,
%	Nanjing University, Nanjing, 210093, China}
%\affiliation{
%	Zhejiang University of Water Resources and Electric Power, Hangzhou 310018, China.
%}

\author{Lin Yang}
\affiliation{National Laboratory of Solid State Microstructures \& School of Physics,
	Nanjing University, Nanjing, 210093, China}

%\author{Yuan-Chun Liu}
%\affiliation{National Laboratory of Solid State Microstructures \& School of Physics,
%	Nanjing University, Nanjing, 210093, China}
	
\author{Da Wang}
\affiliation{National Laboratory of Solid State Microstructures \& School of Physics,
	Nanjing University, Nanjing, 210093, China}

\author{Fu-Chun Zhang}
\affiliation{Kavli Institute for Theoretical Sciences \& CAS Center for Excellence in Topological Quantum Computation, University of Chinese Academy of Sciences, Beijing 100190, China}
\affiliation{Collaborative Innovation Center of Advanced Microstructures, Nanjing University, Nanjing 210093, China}

\author{Qiang-Hua Wang}
\email{qhwang@nju.edu.cn}
\affiliation{National Laboratory of Solid State Microstructures \& School of Physics,
	Nanjing University, Nanjing, 210093, China}
\affiliation{Collaborative Innovation Center of Advanced Microstructures, Nanjing University, Nanjing 210093, China}

\date{\today}% It is always \today, today,
	%  but any date may be explicitly specified
	
\begin{abstract}
We investigate the twisted bilayer graphene by a two-orbital Hubbard model on the honeycomb lattice. The model is studied near 1/4 band filling by using the singular-mode functional renormalization group theory. Spin-triplet $f$-wave pairing is found from weak to moderate coupling limit of the local interactions, and is associated with the Hund's rule coupling and  incommensurate spin fluctuations at moderate momenta.
\end{abstract}
	
\pacs{74.20.-z, 71.27.+a, 74.20.Rp}
	
    %\pacs{74.20.-z}{Theories and models of superconducting state}
	%\pacs{74.20.Pq}{Electronic structure calculations}
	%\pacs{74.20.Rp}{Pairing symmetries (other than s-wave)}
	%\pacs{: 74.20.-z, 74.25.Jb, 74.70.Dd}
	%74.70.Xa Pnictides and chalcogenides
	%75.30.Fv  Spin-density waves
	%74.70.Wz  Carbon-based superconductors
	%81.05.ue  Graphene
	%73.22.Pr  Electronic structure of graphene
	%74.20.Rp  Pairing symmetries (other than s-wave)
	%74.20.-z  Theories and models of superconducting state
	%71.27.+a  Strongly correlated electron systems; heavy Fermions
	%64.60.ae  Renormalization-group theory
	%71.10.-w Theories and models of many-electron systems
	%74.62.Dh Effects of cystal defects, doping and substitution (Superconductivity)
	%74.20.Mn : Nonconventional mechanisms
	%74.25.Dw : Superconductivity phase diagrams
	%74.70.-b :  Superconducting materials other than cuprates
	%(for cuprates, see 74.72.-h; for superconducting films, see 74.78.-w)
	
\maketitle
	
\section{Introduction}
Recently, there is considerable interest in twisted bilayer graphenes (TBG). When two layers of graphenes are stacked and mutually twisted by a specific small angle, periodic Moir\'e pattern appears. The unitcell can be enlarged significantly (with respect to that in the parent graphene), containing tens of thousands atoms. Near the charge neutral point (CNP), the low energy electronic states are mainly derived from those near the parent Dirac points, scattered and recombined by the interlayer coupling. Amazingly, at some magic twisting angles, the four low-energy bands (near the Fermi level) become essentially flat and detached from higher energy ones. \cite{TBG_band1, TBG_band2, TBG_band3, TBG_band4, TBG_band5, TBG_band6} The dispersion remains linear near the CNP, but doping becomes much easier since filling of the entire set of mini flat bands amounts to adding just 4 electrons per unitcell in the superlattice. The effect of interactions becomes important. By the uncertainty principle the kinetic energy scales as $1/l^2$, where $l$ is the linear size of the unitcell, while the long-range Coulomb interaction scales as $1/l$ and can overwhelm the kinetic energy as $l$ becomes large. Dielectric screening from higher-energy bands (above or below the Fermi level) can make the Coulomb interaction short-ranged and consequently also scale as $1/l^2$. Even under this circumstance, the effect of interaction becomes important as the density of states (DOS) at the Fermi level becomes large. Indeed, recent experiment on TBG reveals Mott-like insulating states when the mini flat bands are 1/4 and 3/4 filled.\cite{caoyuan1, tuning_SC_TBG} The conductance in magnetic field indicates that the insulating state is spin-unpolarized.\cite{tuning_SC_TBG} More interestingly, superconductivity (SC) is observed slightly away from 1/4 filling. \cite{caoyuan2} The SC transition temperature $T_c \sim  1.7$ K, and the Fermi energy $E_F\sim 10$ meV in the lower narrow bands. The ratio $T_c/E_F$ is even higher than that in high $T_c$ cuprates, suggesting that TBG can also be taken as a high-$T_c$ superconductor, and the SC therein is very likely unconventional. More recently, SC is also observed near 3/4 filling, but $T_c$ is much lower.\cite{tuning_SC_TBG}

Theoretical concensus is not yet reached regarding the mechanisms underlying the insulating and SC states.
Since the insulating gap at 1/4 and 3/4 fillings in the experiment is about one order of magnitude smaller than the width of the mini bands, the gap may either come from symmetry breaking, \cite{origin_of_mott_and_sc}
or from correlation effects at the verge of the Mott limit.\cite{Josephson_Lattice_TBG, Wigner_Crystallization_TBG, insulating_state_TBG, VB_extended_Hubbard, Charge-transfer_insulation, Moire_band_TBG}
The SC state has recently been discussed in terms of correlation effects \cite{Unconventional_SC_TBG, SC_from_Valley_Fluctuations, Novel_phases_TBG, Topological_SC_TBG, Pairing_symmetry_TBG, d+id_SC_TBG, SC_and_CDW_TBG, FRG_TBG, KL_SC, AFM_d_wave, KL_SC2, TSC_TBG, MFT_TBG, vortex-antivortex_TBG} as well as electron-phonon coupling. \cite{phonon-mediated_SC, phonon_driven_TBG} There are also hot discussions on the appropriate effective lattice model for the mini flat bands. Since the density of low energy states concentrates on the AA stacking positions,\cite{TBG_band2} a two-orbital Hubbard model on a triangular lattice is proposed.\cite{d+id_SC_TBG}  Alternatively, there are arguments, in view of the band degeneracy at the CNP, in favor of a two-orbital model on an effective honeycomb lattice.\cite{A_Model_for_TBG, extended_Hubbard_model_for_TBG,origin_of_mott_and_sc} Depending on the normal state band structure (and the type of interactions), the proposed pairing symmetry in the SC state varies from $d+id'$-, $p+ip'$- to more conventional $s$-wave. 

Here we study a two-orbital Hubbard model on the honeycomb described by Eq.\ref{H0} below. The model is similar to that proposed in Refs.[\onlinecite{A_Model_for_TBG},\onlinecite{extended_Hubbard_model_for_TBG}]. The difference is the particle-hole asymmetry is introduced, which is known to be present and causes asymmetric behaviors at 1/4 and 3/4 fillings.\cite{caoyuan1}
We limit ourselves to filling levels near 1/4 filling, where SC is experimentally found to be much stronger than that near 3/4 filling.\cite{caoyuan2, tuning_SC_TBG} We investigate various electronic instabilities on equal footing by using the singular-mode functional renormalization group (SM-FRG).
\cite{Triplet_pairing_and_topoSC, negative_isotope_effect, TSC_FM, AFM_f-wave_and_chiral_f-wave_SC, FRG_and_VMC_for_graphene, BC3} Here we will not address the Mottness at 1/4 filling, which is beyond the realm of FRG that requires a metalic normal state as the starting point.  

Our main result are as follows. For quite general local interactions, and from weak coupling to moderate coupling, we find robust $f$-wave SC, related to incommensurate spin fluctuations at moderate momenta. The pairing function describes local orbital-singlet and spin-triplet Cooper pairs. 

The remainder of this paper is structured as follows. In Sec.\ref{sec:mm} we specify the model. In Sec.\ref{sec:frg} we investigate the electronic instabilities for moderately strong interactions by SM-FRG. In Sec.\ref{sec:weak} we use the weak coupling theory to study the model and compare the results with FRG theory. Finally, in Sec.\ref{sec:summary} we provide the conclusion and further remarks.

\section{Model}\label{sec:mm}

\begin{figure}
	\includegraphics[width=0.9\columnwidth]{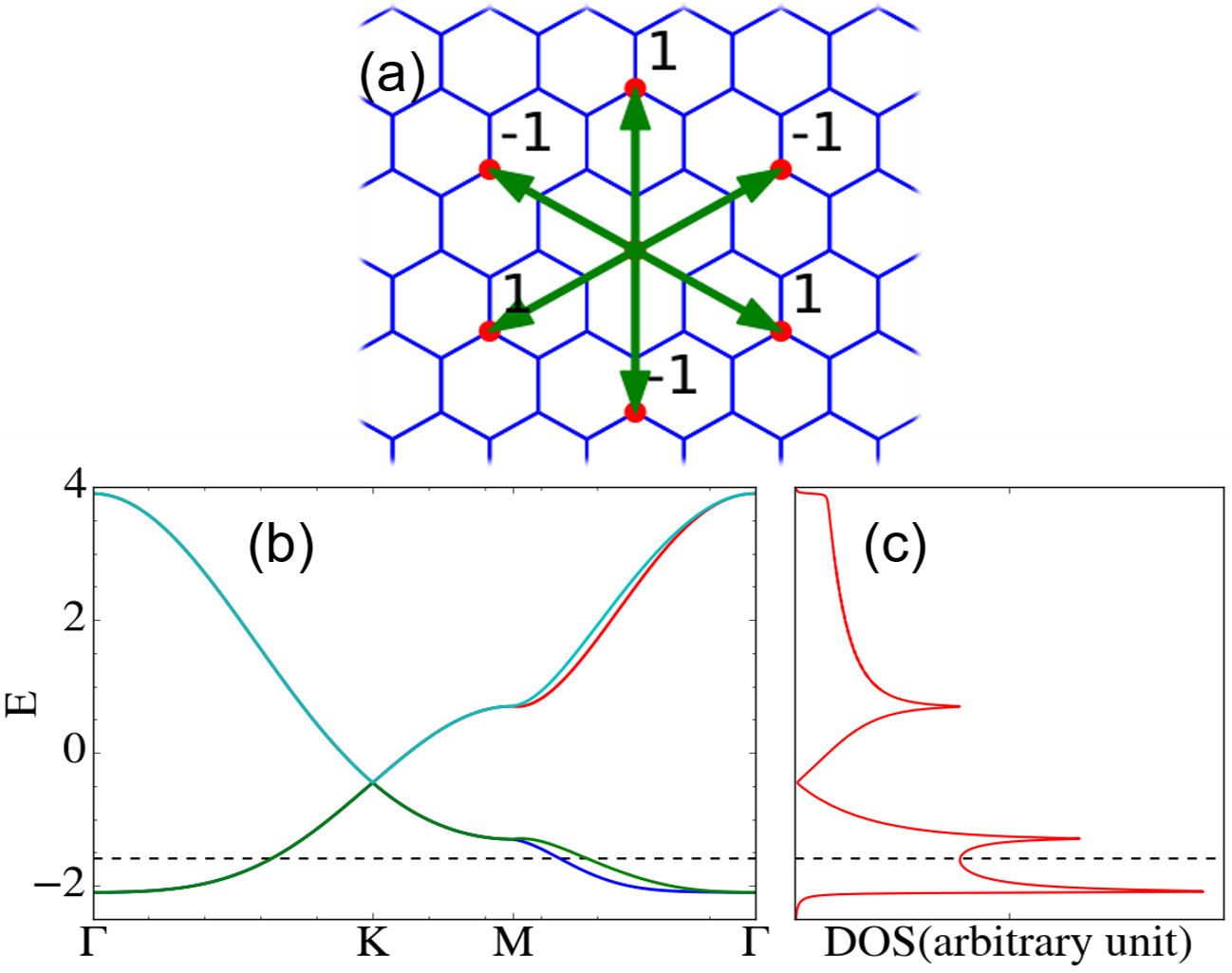}
	\caption{(a) The structure of honeycomb lattice. The green arrows show fifth-neighbor bonds, with the $f$-wave sign $f_{\<ij\>}$ for the hexagonal warping term. (b) Band dispersion along high-symmetry cuts in the Brillouine zone. The momenta $\Ga$, $K$ and $M$ can be found in the left inset of Fig.\ref{frg1}. (c) The normal state DOS. The horizontal dashed lines in (b) and (c) highlight the Fermi level at 1/4 band filling, or electron density $n_e=1$ (per site).} \label{dispersion}
\end{figure}

The TBG is $D_{3}$ symmetric. An effective free Hamiltonian describing low energy electrons on the Moir\'e superlattice can be written as,\cite{A_Model_for_TBG, extended_Hubbard_model_for_TBG}
\eqa
H_0 = &&-\mu\sum_{i}\psi_i^\dag \psi_i + \sum_{m=1,2}\sum_{\<ij\>\in N_m} t_m\psi_i^\dag \psi_j \nn
&&+t_5'\sum_{\<ij\>\in N_5} \psi_i^\dag i\tau_2 \psi_j f_{\<ij\>}.\label{H0}
\eea
Here $\mu$ is the chemical potential, $\psi_i$ is a spinor composed of $c_{ia\si}$ annihilating an electron at site $i$ (the AB or BA stacking position on the Moir\'e superlattice) on orbital $a\in (1, 2)$ with spin $\si\in (\ua,\da)$,  $N_{1,2,5}$ denotes the (first, second, and fifth) neighbor on all directions, $i\tau_2$ is the $2\times 2$ antisymmetric tensor in the orbital basis. The $N_5$ bonds are shown in Fig.\ref{dispersion}(a), together with the $f$-wave factor $f_{\<ij\>}=\pm 1$. We set $t_2 = 0.15t_1$ and $t'_5 = -0.02t_1$, and use $t_1=1$ (corresponding roughly  to 1 meV) as the unit of energy henceforth. Note $H_0$ is invariant under spin-SU(2) and time reversal (TR). It can also be endowed with inversion symmetry if the inversion operator also flips the sublattices and orbitals, see Appendix. These symmetries make spin-singlet and spin-triplet Cooper pairs sharply defined. Fig.\ref{dispersion} shows the band dispersion along high symmetry cuts (b) and the normal state DOS (c), in qualitative agreement to that in Refs.[\onlinecite{TBG_band6}, \onlinecite{caoyuan2}]. The CNP is at the band touching point $K$, and corresponds to 1/2 filling of the entire set of bands. The dispersion near $K$ remains linear, and the $K$ point is 4-fold degenerate (aside from spin). Note the two-fold band degeneracy along $\Ga$-$K$-$M$, which is preserved by the $t_5'$-warping. Replacing the fifth-neighbor bonds by shorter ones breaks the above degeneracy. On the other hand, a first-neighbor inter-orbital hopping $t_1'$ (not included here) would also break this degeneracy, resulting in quasi-nested Fermi surface near 1/4 filling.\cite{d+id_SC_TBG}

\begin{figure}
	\includegraphics[width=0.7\columnwidth]{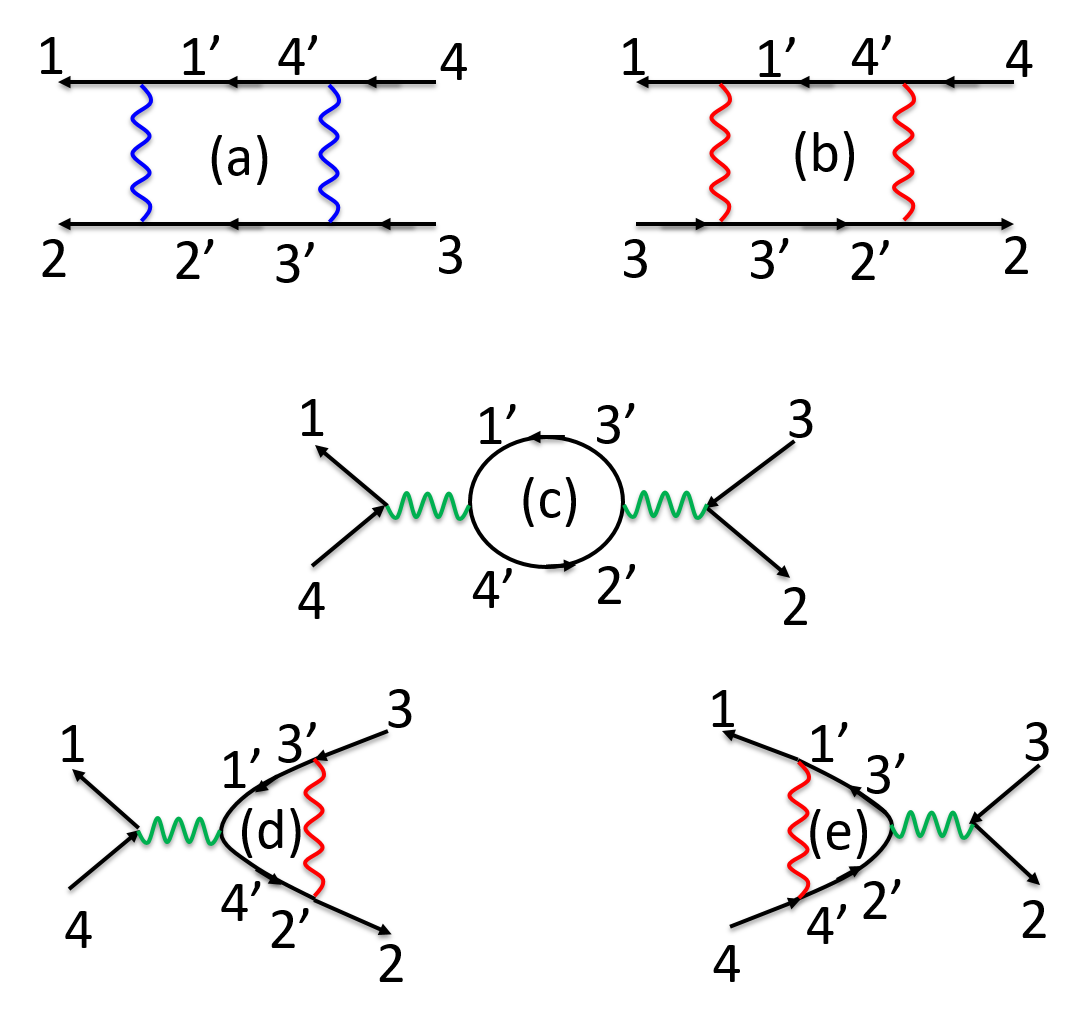}
	\caption{One-loop diagrams contributing to $\p\Ga_{1234}/\p\La$, quadratic in $\Ga$ itself (wavy lines, with incoming or outgoing fermions labelled by the numerical indices). The spin is conserved along fermion lines. The color of the wavy line indicates the scattering of fermion bilinears in the pairing (blue), crossing (red) and direct (green) channels. The loop integrations are to be differentiated with respect to the running scale $\La$ that regulates the single-particle propagators.}\label{fig:feynman}
\end{figure}

Note that the 1/4 band filling corresponds to one electron per site (or two electrons per unitcell). For clarity, we will henceforth use the electron density $n_e$ (per site) to reflect the filling level, with the understanding that $n_e>1$ ($n_e<1$) means electron (hole) doping away from 1/4 band filling.  

The interactions between the electrons are assumed local, 
\eqa H_{I}  &= &U \sum_{ia} n_{ia\ua}n_{ia\da}+J \sum_{i,a>b,\si\si'}c^{\dag}_{ia \si}c_{ib \si} c^{\dag}_{ib\si'}c_{ia\si'} \nn
&+&U' \sum_{i,a>b} n_{ia}n_{ib}+J' \sum_{i,a\neq b}c^{\dag}_{ia\ua}c^{\dag}_{ia\da} c_{ib\da}c_{ib\ua}, \label{H_I} \eea
where $n_{ia}=\sum_{\si} n_{ia\si} = \sum_{\si} c_{ia\si}^\dag c_{ia\si}$, $U$ is the intra-orbital repulsion, $U'$ is the inter-orbital repulsion, $J$ is Hund's rule coupling, and $J'$ is the pair hopping term. We assume the Kanamori relations $U=U'+2J$ and $J'=J$ to take $(U',J)$ as independent parameters, although such relations are exact only in the case of rotationally invariant atomic limit.

\section{SM-FRG results} \label{sec:frg}

\begin{figure}
	\includegraphics[width=0.8\columnwidth]{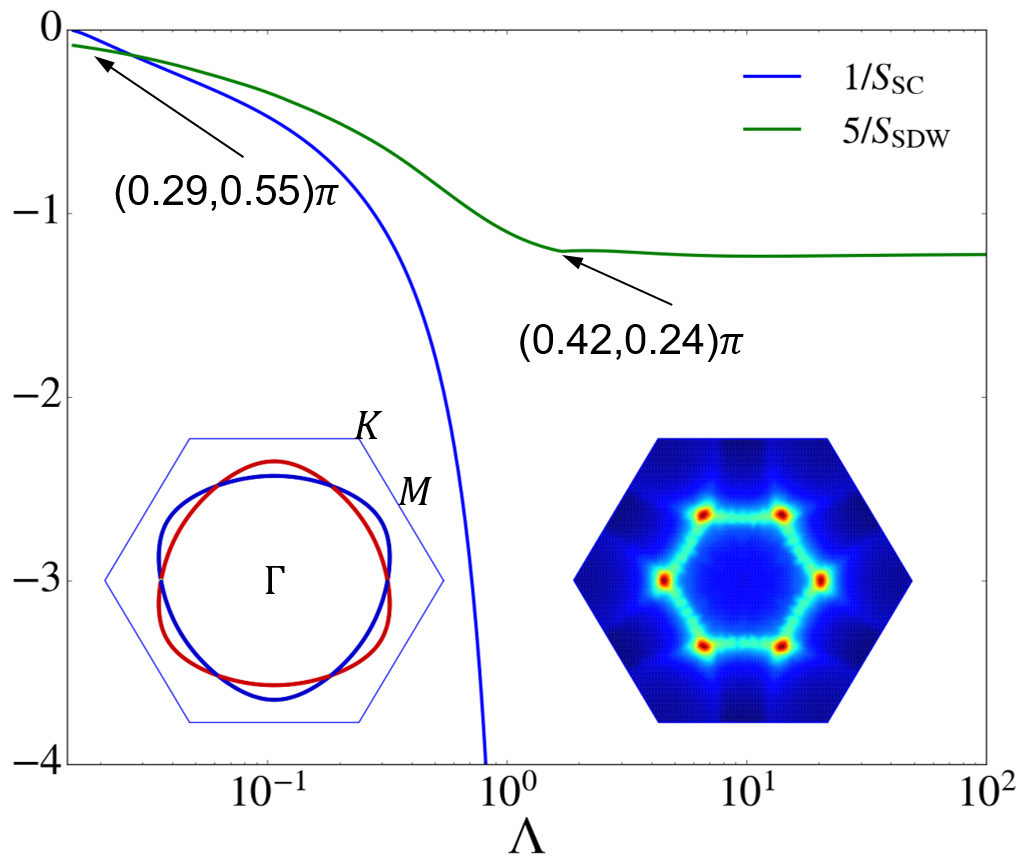}
	\caption{FRG flow of the most negative eigenvalue $S_{\rm SC,SDW}$ versus the running energy cutoff $\La$ for $(U',J)=(2.5,0.8)$ and  $n_e=1.106$. The texted arrows indicate the collective momentum $\v Q$, shown as $\v Q/\pi$, associated with $S_{\rm SDW}$. The collective momentum in the SC channel is fixed at $\v Q=0$. The left inset shows the FRG-derived pairing function on the Fermi surface, which is $\pm 1$ where the color is red/blue, up to a global factor. The outer hexagon is the Brillouine zone, with high symmetry points indicated. The right inset shows $-S_{\rm SDW}\equiv |S_{\rm SDW}|$ in the momentum space. The signal is strong (weak) where the color is red (deep blue). } \label{frg1}
\end{figure}

Here we treat the correlation effect by SM-FRG. The idea of FRG\cite{Wetterich_effective_potential} is to obtain the one-particle-irreducible (1PI) 4-point interaction vertices $\Ga_{1234}$ (where numerical index labels the single-particle state) for quasi-particles above a running infrared energy cut off $\La$ (which we take as the lower limit of the continuous Matsubara frequency). Starting from $\La=\infty$ where $\Ga$ is specified by the bare parameters in $H_I$, the contribution to the flow (toward decreasing $\La$) of the vertex, $\p\Ga_{1234}/\p\La$, is illustrated in Fig.\ref{fig:feynman}. At each stage of the flow, we decompose $\Ga$ in terms of eigen scattering modes (separately) in the SC, SDW and CDW channels to find the negative leading eigenvalue (NLE). Notice that the NLE is a function of the collective momentum. The divergence of the most netative eigenvalue (MNE) signals an emerging order at the associated collective momentum, with the internal microscopic structure described by the eigenfunction. The technical details can be found elsewhere,\cite{Triplet_pairing_and_topoSC, negative_isotope_effect, TSC_FM, AFM_f-wave_and_chiral_f-wave_SC, FRG_and_VMC_for_graphene, BC3} and also in the self-contented Appendix.

Fig.\ref{frg1} shows the RG flow of the MNE $S_{\rm SC,SDW}$ in the SC and SDW channels, for $n_e=1.106$ and $(U',J)=(2.5,0.8)$. The CDW channel is weak during the entire flow and not shown here. We see the SDW channel dominates in the high energy window. The collective momentum associated with the MNE evolves with decreasing $\La$ (see the arrows). Attractive MNE in the SC channel (with collective momentum $\v Q\equiv 0$) emerges as $\La<1$, where the SDW channel is also enhanced. This is a manifestation of the Luttinger-Kohn mechanism, namely, fluctuations in the particle-hole (PH) channel have projections in the particle-particle (PP) channel. FRG makes this notion even sharper, namely, it is those enhanced PH fluctuations (as the energy scale is lowered) that is related to (or contribute to) attractive pair interactions. At lower energy scales, the SC channel flows faster and diverges eventually at $\La_c\sim 1.55\times 10^{-2}$. From our SM-FRG, the associated pairing function (the MNE scattering mode in the SC channel) is $i\tau_2$ in the orbital basis. This describes pairing between two electrons in an orbital-triplet. By fermion antisymmetry the spin part must be a triplet. Such a pairing is favored by the local Hund's rule coupling in Eq.\ref{H_I}. We will come back to this point at the end of this section. The pairing function projected in the band basis is shown on the Fermi surface (FS) in the left inset, which is $\pm 1$ on the red/blue segment of the FS (up to a global factor), showing $f$-wave symmetry. Note the $f$-wave form belongs to the A$_{1}$ representaiton of $D_3$, invariant under both $C_3$ and $C_2'$. Further discussions on the pairing symmetry can be found in the Appendix.

\begin{figure}
	\centering
	\includegraphics[width=0.95\columnwidth]{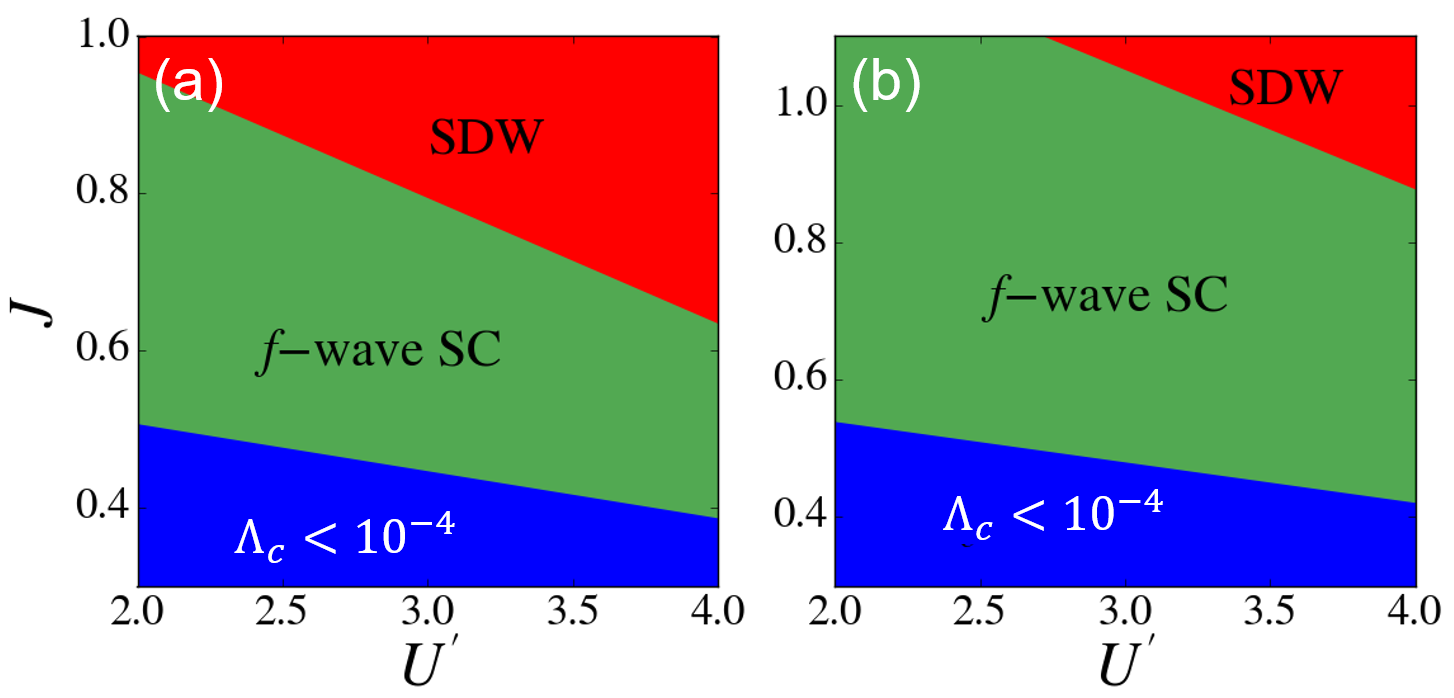}
	\caption{The phase diagram in the $(U',J)$ parameter space. The electron density is (a) $n_e=1.106$, and (b) $n_e=0.902$. The divergence scale $\La_c<10^{-4}$ in the blue regime.} \label{phase}
\end{figure}

Several remarks are in order. First, the local pairing is insensitive to FS nesting. Indeed, similar results are obtained for $(t_2,t_5') = (0.12, -0.036)$, although better quasi-nesting on the FS arises. Second, the pairing function in the orbital basis, $i\tau_2$, is robust even if we get rid of the warping term. Third, the crossing of FS pockets is protected by the $D_3$ symmetry, but breaking of $D_3$ would generate anti-crossing and gap nodes. Spin-orbital coupling may also generate gap nodes, but is expected to be weak in TBG. Finally, from the right inset in Fig.\ref{frg1}, we see that at the final stage of the RG flow, the NLE $S_{\rm SDW}$ in the momentum space peaks roughly at the mid point on $\Ga$-$K$. This means that our $f$-wave triplet is tied to incommensurate spin fluctuations at moderate momenta, instead of the usual small-momentum spin fluctuations as in the case of Sr$_2$RuO$_4$. \cite{epl_wang,uniaxial,biaxial,nearnode}

In Fig.\ref{phase} we present the phase diagram in the parameter space, for electron density $n_e=1.106$ (a) and $n_e=0.902$ (b). These may be understood as electron and hole doping away from 1/4 band filling. In both cases of electron and hole doing, $f$-wave pairing is observed for intermediate values of $J$ with sizable divergence scale $\La_c$, but as $J$ increases further, the system enters the SDW phase. We note that local triplet pairing is favorable even at the mean field level if $J>U/3$ for two-orbital models. \cite{TRI_SC,Zhang} This condition is further relaxed in Fig.\ref{phase}, since the charge screening effect captured by FRG makes the ratio $J/U$ effectively larger. \cite{Yang} In agreement with the experiment, SC arises from both electron and hole doping away from the Mott limit. 

\section{Weak coupling limit} \label{sec:weak}
In this section we investigate the weak coupling limit of the model. On one hand, further approximations to FRG can be made, providing clearer understanding of the pairing mechanism. On the other hand, consistency from weak to moderate coupling, if any, provides evidence of the robustness of the $f$-wave SC state.

\begin{figure}
	\centering
	\includegraphics[width=\columnwidth,clip]{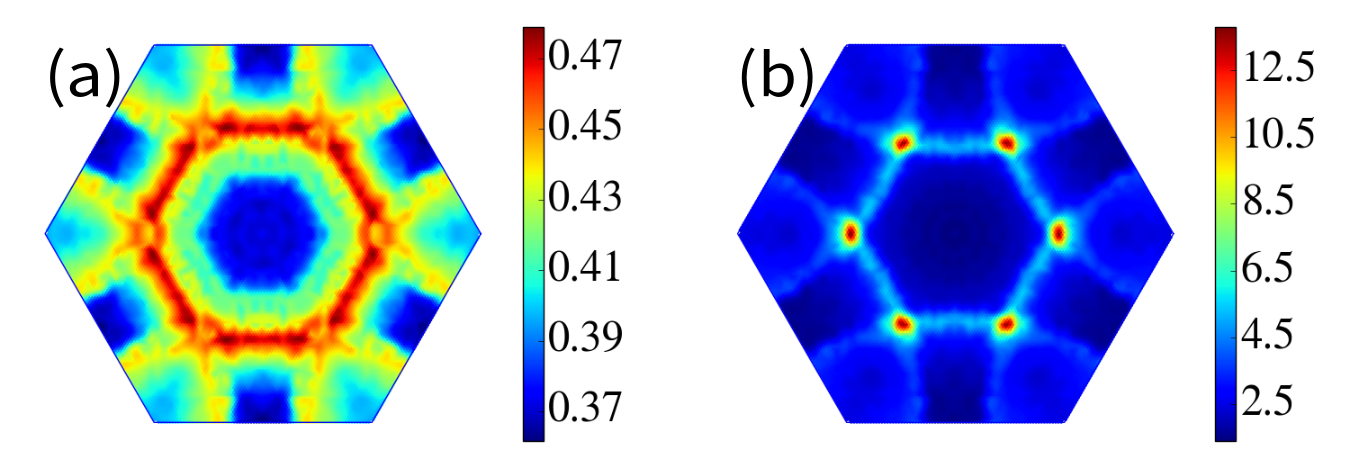}
	\caption{The leading positive eigenvalue of the spin susceptibility matrix (in the orbital basis) in the momentum space, in the weak coupling theory, for $n_e=1.106$ and $(U', J)=(2.5,1)/g$, with the artificial scaling factor $g=2.6$ to avoid divergence in RPA. (a) Bare susceptibility. (b) RPA-renormalized spin susceptibility. There are peaks at six momenta $\v Q_{1,\cdots,6}$, with $\v Q_1=(0.625\pi,0)$.} \label{xs}
\end{figure}

In the weak coupling limit, we may ignore the mutual overlaps in the SC/SDW/CDW channels. Then each channel flows independently, and can be solved exactly, see the Appendix. This corresponds to the ladder approximation in the SC channel, and random phase approximation (RPA) in the DW channels. Eventually, we project interactions in all channels onto the effective pairing interaction $V(\v k,\v k')$ on the Fermi surface (with proper elimination of over counting), and solve the following Eliashberg equation
\eqa -\oint \frac{dl'}{(2\pi)^2 v(l')} V(l, l') \Del(l') = \la \Del(l), \label{eq:bcs}\eea 
to get the leading pairing function. Here $l$ is the momentum path on the Fermi pockets, $v$ is the Fermi velocity, and the integration sums implicitly over Fermi pockets. The eigenvalue $\la$, or the coupling constant, is to be related to the transition temperature as $T_c\sim 1.14\La_0 e^{-1/\la}$, where $\La_0$ is the energy scale (the temperature, e.g.) at which the ladder/RPA is performed. This mechanism is referred to as fluctuation-exchange (FLEX),\cite{FLEX} emphasizing the role of DW fluctuations in triggering superconductivity for repulsive models. More technical details of the FLEX for our case can be found in Appendix. Since divergence occurs too soon in RPA, this approach is better justified in the weak coupling limit.

\begin{figure}
	\includegraphics[width=0.9\columnwidth]{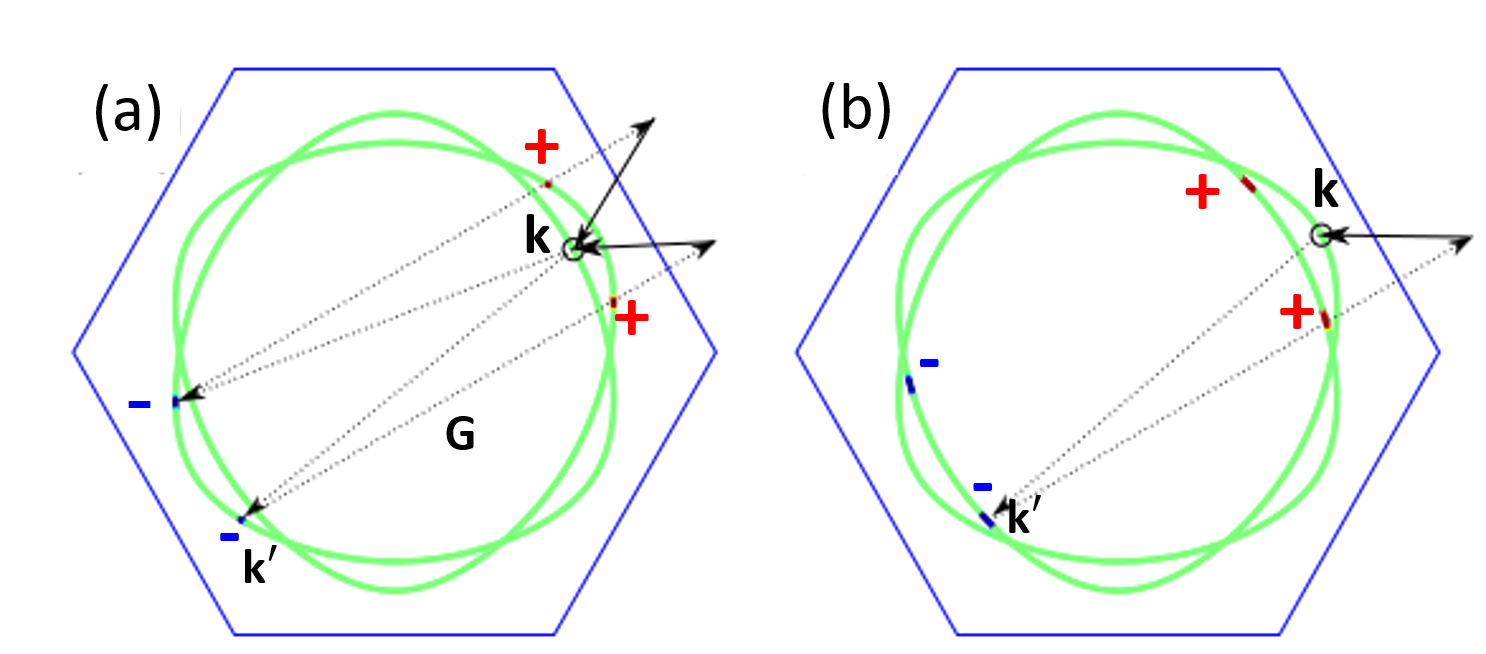}
	\caption{The pair interaction in the band basis $V(\v k,\v k')$ as a function of $\v k'$ (on the FS), for a fixed $\v k$ (open circle) on the (a) inner pocket, and (b) outer pocket. Here $V(\v k, \v k')$ is constructed from contributions by spin fulcutaions at momenta $\v Q_i$'s, hence is nonzero only on some discrete points of $\v k'$, with the sign indicated by $\pm$. The solid arrows represent $\v Q_i$. The geometry shows $\v k'-\v k\pm \v Q_i=\v G$ where $\v G$ is a reciprocal vector.} \label{vkk}
\end{figure}

For comparison, we scale down the `strong' bare interactions, in the same order as used in the previous section, by a factor of $g=2.6$. For $n_e=1.106$ and $(U',J)=(2.5,1)/g$, we find the resulting pairing function is identical to that shown in the left inset of Fig.\ref{frg1}. To understand this result, we present the spin susceptibility in Fig.\ref{xs}. Compared to the bare susceptibility in (a), the renormalized one in (b) is more concentrated at the six momenta $\v Q_{1,\cdots,6}$. This pattern is closely similar to that of the NLE in the SDW channel shown in Fig.\ref{frg1} (right inset), and the associated spin fluctuations may trigger the $f$-wave pairing according to FLEX. To single out such an effect, we consider the contributions to $V(\v k,\v k')$ from  the SDW channel with collective momentum $\v q\sim \v Q_{i=1,\cdots,6}$. To filter away the interactions irrelevant to odd-parity pairing, we antisymmetrize $V(\v k,\v k')$ with respect to $\v k\ra -\v k'$ and/or $\v k'\ra -\v k'$. The resulting pair interaction $V(\v k,\v k')$ is shown in Fig.\ref{vkk} for $\v k$ on the inner pocket (a) and outer pocket (b). In each case, we can find two positions of $\v k'$ with $V(\v k,\v k')<0$, satisfying $\v Q_i+\v k'-\v k = \v G$ for a specific $\v Q_i$, and $\v G$ is a reciprocal vector. The pairing function should be of the same sign on $\v k$ and $\v k'$ if $V(\v k,\v k')<0$, and vice versa. Inspection of Fig.\ref{vkk} shows that this requires the nearby inner and outer pockets to be antiphase. This sign structure is in full agreement with the $f$-wave pattern discussed in the previous section, showing spin fluctuations at (and near) the momenta $\v Q_i$'s tend to trigger the triplet $f$-wave pairing. 

We should point out, however, in the above approximation, the $f$- and $d$-wave pairing are close in eigenvalue $\la$, and the leading one becomes $d$-wave if $J$ is decreased, e.g., $(U, J)=(2.5,0.6)/g$. This is not the case in the full-fledged SM-FRG for the unscaled interactions (or $g=1$), where $f$-wave is robust, see Fig.\ref{phase}.

\section{Summary and discussion}\label{sec:summary}
We have applied functional renormalization group and weak coupling theory to study the superconductivity in a two-band Hubbard model of TBG near 1/4 filling. The pairing function is found to be $f$-wave spin-triplet on both sides of 1/4 filling, from weak to moderate coupling limit of the interactions. The pairing mechanism of the $f$-wave triplet is due to effective attraction between two electrons in the spin-triplet and orbital-singlet state on the same site. In the weak coupling scenario, the SC is related to incommensurate spin fluctuations at moderate momenta. 

We remark that our $f$-wave pairing is time-reversal invariant, in contrast to $p+ip'$ or $d+id'$ in previous studies. \cite{Unconventional_SC_TBG, Topological_SC_TBG} In experiment, the difference can be easily distinguished by $\mu$SR. In theory, the difference has much to do with the starting model. For example, if we include inter-orbital hopping on first-neighbor bonds as in Ref.\onlinecite{d+id_SC_TBG}, our FRG also yields $d+id'$-wave SC. Therefore, more accurate understanding of the normal state band structure is necessary. On the other hand, we have also studied SC near 3/4 filling, but the critical scale is much smaller than that near 1/4 filling. This could be understood from the smaller DOS near 3/4 filling, and is consistent with the experiment. Finally, the way how the $f$-wave SC and the incommensurate SDW fluctuations evolve into the Mott insulating state at 1/4 filling is an interesting but open topic.

\acknowledgments{QKT thanks Wei-Cheng Bao for technical helps, and thanks Yuan-Chun Liu for discussions. The project was supported by the National Key Research and Development Program of China (under Grant No. 2016YFA0300401), the National Basic Research Program of China by MOST (under Grant No. 2014CB921203), and the National Natural Science Foundation of China (under Grant Nos.11574134, 11504164 and 11404383). FCZ also acknowledges the support by the Strategic Priority Research Program of the Chinese Academy of Sciences (under Grant No. XDB28000000).}\\

\section{Appendix}\label{appendix}

%\beginsupplement
\subsection{Symmetry of the normal state and pairing function}

The single-particle part of $H_0$ can be written as, for both spin species and at momentum $\v k$,
\eqa h_\v k = \eps_0(\v k) + \eps_1(\v k) s_1 + \eps_2(\v k) s_2 + \eps_3(\v k)\tau_2,\eea
where $s_i$'s are Pauli matrices in sublattice basis, identity matrices are dropped without causing any ambiguity, and
\eqa &&\eps_0(\v k)=-\mu + t_2\sum_{\v b\in N_2} \cos \v k\cdot \v b,\\
     &&\eps_1(\v k) = t_1 \sum_{\v b\in N_{1A}}\cos\v k\cdot\v b,\\
     &&\eps_2(\v k)= -t_1\sum_{\v b\in N_{1A}}\sin\v k\cdot\v b,\\
     &&\eps_3(\v k)=-t_5'\sum_{\v b\in N_5} f_\v b \sin\v k\cdot\v b. \eea
Here $N_{1A}$ denotes the set of three first-neighbor bonds radiating from a site on the A-sublattice of the honeycomb lattice, $N_2$ is the set of six second-neighbor bonds, $N_5$ is the set of six fifth-neighbor bonds, and $f_\v b$ is the $f$-wave sign factor illustrated in Fig.\ref{dispersion}(a).   
Note that $\eps_{0,1}(\v k)$ is even, and $\eps_{2,3}(\v k)$ are odd, under $\v k\ra -\v k$. In fact $\eps_3(\v k)$ transforms as $f$-wave in $\v k$ a la the sign structure of $f_\v b$. So $h_{-\v k}\neq h_{\v k}$, breaking the inversion symmetry in a naive manner. However, we can endow the inversion symmetry for $h_\v k$ by combining the actions
\eqa s_i \ra s_1 s_i s_1, \ \ \ \tau_i \ra \tau_1 \tau_i \tau_1.\label{eq:hk} \eea
The first operation means flip of sublattices upon inversion, defining the symmetry center at the center of the holo hexagon in real space. The second operation means flip of orbitals. (The action on the orbital by inversion can be generalized to any action that causes $\tau_2\ra -\tau_2$, but here we stick to orbital flip for definiteness.) In this definition, the inversion operator $\cP$ acts on $h_\v k$ as 
\eqa \cP h_\v k \cP^{-1} = s_1 \tau_1 h_{-\v k} s_1 \tau_1 = h_{\v k}.\label{eq:inversion}\eea
So inversion relates $h_{\v k}$ and $h_{-\v k}$ in a nontrivial way. 
In addition, the TR invariance can be expressed as
\eqa \cT h_{\v k} \cT^{-1} = h_{-\v k}^*=h_\v k.\label{eq:TR}\eea  
Note we defined the TR operator $\cT$ irrespectively of spin. This is possible because the spin-SU(2) symmetry enables us to treat spins separately. Finally, $h_{\v k}$ is symmetric under $D_3$ acting on $\v k$ alone. 

The band dispersion can be obtained straightforwardly, 
\eqa E^{\pm}(\v k\nu) = \eps_0(\v k) \pm \sqrt{\eps_1^2(\v k)+\eps_2^2(\v k)} + \nu |\eps_3(\v k)|,\eea
where $\nu=\pm 1$ is the eigenvalue of $\tau_2 f_\v k$, with $f_\v k= \sign[\eps_3(\v k)]$. In this band labeling scheme, the band label, say $\nu=1$, may correspond to the positive or negative eigenvalue of $\tau_2$, depending on the sign of $\eps_3(\v k)$. The advantage of the this $\v k$-dependent band labelling is the band energy is explicitly inversion symmetric. It is also the natural labeling scheme according to the ordering of band energies at the same momentum.

The Fermi level cuts the lower two bands described by $E^-(\v k\nu)$. We now discuss these energy bands in more details, and we drop the superscript on $E$ for brevity. For a band state $|\v k\nu\>$ satisfying
\eqa h_\v k |\v k\nu\> = E(\v k\nu)|\v k\nu\>,\eea
we have
\eqa h_{-\v k} s_1\tau_1 |\v k\nu\> = E(\v k\nu) s_1\tau_1 |\v k\nu\>, \eea by inversion symmetry defined in Eq.\ref{eq:inversion}. This requires $s_1\tau_1 |\v k\nu\>$ to be the eigenstate of $h_{-\v k}$ with energy $E(\v k\nu)$, which is identical to $E(-\v k\nu)$ in our band labelling scheme. In other words, 
\eqa \cP |\v k\nu\> = |-\v k\nu\>, \eea up to a phase. 
On the other hand, by TR symmetry defined in Eq.\ref{eq:TR}, 
\eqa h_{-\v k} K|\v k\nu\> = E(\v k\nu) K|\v k\nu\>,\eea where $K$ stands for complex conjugation. This implies 
\eqa K|\v k\nu\> = |-\v k\nu\>, \eea up to a phase.

The full matrix pairing function for the model studied in the main text can be written as $\phi(\v k)\sim i\tau_2$ in the orbital basis. (The spin component will be specified shortly.) Under inversion, $\cP \phi(\v k)\cP^{-1} = s_1\tau_1 \phi(-\v k) s_1\tau_1 = -\phi(\v k)$, so the pairing function is odd under inversion. To see the resulting gap structure more transparently, we project the pairing matrix onto the normal state band basis. We first recall that the field operators in the band and orbital bases are related as
\eqa d_{\v k\nu}^\dag = \sum_m\psi_{\v k m}^\dag \<m|\v k\nu\>, \ \ \ \psi_{\v km}^\dag = \sum_\nu d_{\v k\nu}^\dag \<\v k\nu|m\>,\nonumber\eea
where $m$ donotes a component of $\psi^\dag_{\v k}$ or $|\v k\nu\>$. It is now straightforward to make the transformation
\eqa \psi_\v k^\dag \phi(\v k) \psi_{-\v k}^{\dag,t} \ra d_\v k^\dag \Del(\v k) d_{-\v k}^{\dag,t},\eea
where $t$ means transpose, and the matrix element of $\Del(\v k)$ is given by, 
\eqa \Del^{\nu\nu'}(\v k) = \<\v k\nu |\phi(\v k) K|-\v k\nu'\> = i\nu \del_{\nu\nu'} f_\v k,\eea
which is explicitly odd in $\v k$ and actually transforms as $f_\v k$. This function is shown in Fig.\ref{frg1} (left inset). 

Finally, we include the spin content in the pairing function. By spin-SU(2) and inversion symmetries, the odd-parity Cooper pair has to be a spin triplet, and the resulting matrix pairing function has to be in the form, for the lower two bands, 
\eqa  \Del(\v k)\ra i \eta_3 f_\v k \v d\cdot \vec \si i\si_2,\eea
where $\eta_3$ is the Pauli matrix in the band basis, $\v d$ is a constant vector, and $\vec \si$ is the Pauli vector in the spin basis. The pairing function is also explicitly TR invariant.

\subsection{FRG flow equation}
The idea of FRG\cite{Wetterich_effective_potential} is to obtain the 1PI 4-point interaction vertices $\Ga_{1234}$, as in 
\eqa H_\Ga=\frac{1}{2}\sum_{1\si,2\si', 3\si', 4\si}\psi_{1\si}^\dag \psi_{2\si'}^\dag \Ga_{1234} \psi_{3\si'} \psi_{4\si},\eea
for quasi-particles above a running infrared energy cut off $\La$ (which we take as the lower limit of the continuous Matsubara frequency). The numerical subscript $1=(\v k, a,s)$ labels (momentum, orbital, sublattice). The spin $\si$ is conserved explicitly in the above form and drops out of $\Ga$ effectively. Momentum conservation is assumed implicitly. Equivalently, $\Ga_{1234}$ may be taken as the effective interactions on quasiparticles below the scale $\La$, in the spirit of pseudopotential. Starting from $\La=\infty$ where $H_\Ga$ is specified by the bare interaction $H_I$, the contribution to the flow (toward decreasing $\La$) of the vertex, $\p\Ga_{1234}/\p\La$, is illustrated in Fig.\ref{fig:feynman} in the main text. The SM-FRG is a realization of FRG in terms of scattering between truncated fermion bilinears that are sufficient to capture the potentially singular scattering modes in the quantum manybody system,\cite{Triplet_pairing_and_topoSC, negative_isotope_effect, TSC_FM, AFM_f-wave_and_chiral_f-wave_SC, FRG_and_VMC_for_graphene, BC3} see below.

Before proceeding, we notice that for a system with featureless Fermi surface(s), standard RG dimension-counting reveals that all four-point interactions are marginal. This means higher-order vertices and the frequency dependence in the 4-point vertex $\Ga_{1234}$ are irrelevant and can be dropped, as long as instabilities occur at low energy scales (to justify the RG argument). In this approximation, the single-particle self-energy correction is frequency independent and can be absorbed in the normal state Hamiltonian. We will therefore concentrate on the flow of $\Ga_{1234}$ only.

It turns out to be useful to view $\Ga$ as scattering matrices for fermion bilinears,
\eqa \Ga_{1234}=P_{12,43}=C_{13,42}=D_{14,32},\label{eq:pcd}\eea
where $P$ is the matrix in the pairing channel, $C$ in the crossing channel, and $D$ in the direct channel.
Then the RG flow equation, shown schematically in Fig.\ref{fig:feynman}, can be written compactly as
\eqa \frac{\p \Ga_{1234}}{\p \La} &&=[(C-D)\chi'_{ph}D]_{14,32}-[P\chi'_{pp}P]_{12,43} \nn &&+[D\chi'_{ph}(C-D)]_{14,32}+ [C\chi'_{ph}C]_{13,42},\label{eq:flow}\eea
where the products within the square brackets are understood as convolution in the bilinear labels, see Fig.\ref{fig:feynman}, and $\chi'_{pp/ph}$ are differential susceptibilities, as matrices in the bilinear basis,
\eqa &&[\chi'_{pp}]_{12, 34} = \frac{\p}{\p\La}\int\frac{d\w \th(|\w|-\La)}{2\pi}G_{13}(i\w) G_{24}(-i\w),\\
	&&[\chi'_{ph}]_{12, 34} = -\frac{\p}{\p\La}\int \frac{d\w \th(|\w|-\La)}{2\pi} G_{13}(i\w) G_{42}(i\w),
	\label{eq:chi} \eea
where $G_{12}(i\w_n)=-\<\psi_1(i\w_n)\psib_2(i\w_n)\>$ is the normal state Matsubara Green's function. In actual calculations, the loop integration in Eq.\ref{eq:flow} is performed in momentum space.

\subsection{Singular scattering modes in collective channels}
For spin SU(2) invariant systems, there are three types of collective scattering channels, namely, SC, SDW, and CDW. The corresponding scattering matrices are given by
\eqa V^{\rm SC} = P,\ \ V^{\rm SDW} = -C, \ \ V^{\rm CDW} = 2D-C. \eea
We will use $(P,C,D)$ and $V^{\rm SC/SDW/CDW}$ interchangeably in the above sense. We now discuss how a diverging or singular eigenmode of the above scattering matrices tells about the emerging order. 
To basic idea is most easily explained by ignoring the spin, orbital and sublattice labels for the moment. Let us consider fermion bilinears limited to a set of relative displacement $\v r$, say $\v r\in (\v r_m, m=1,\cdots )$.
We first rewrite the effective interaction $H_\Ga$ on quasiparticles as, up to an unimportant global factor,
%\begin{widetext}
	\eqa H_\Ga &&\sim \sum \psib_{\v R_0}\psib_{\v R_0+\v r_m} V^{\rm SC}_{mn}(\v R)\psi_{\v R+\v R_0+\v r_n }\psi_{\v R+\v R_0} \nn &&= \sum\psib_{\v k+\v q}\psib_{-\v k}f_m(\v k) V^{\rm SC}_{mn}(\v q) f_n^*(\v k')\psi_{-\v k'}\psi_{\v k'+\v q}.\eea
%\end{widetext}	
Henceforth summation over all repeated indices is implied by a blind $\sum$ for brevity. The first (second) equality is in the real (momentum) space, $V^{\rm SC}_{mn}(\v R)\equiv V^{\rm SC}_{(0, \v r_m), (\v R, \v R+\v r_n)}$, and $f_l(\v k)=e^{i\v k\cdot\v r_l}$ is a basic lattice harmonics, or form factor. The matrix $V^{\rm SC}(\v q)$ is hermitian and can be decomposed as, dropping $\v q$ for brevity,
\eqa V^{\rm SC}_{mn} = \sum_\al \phi_m^\al S_\al \phi_n^{\al*}, \eea
where $\al $ labels the eigenstate $\phi^\al$ with eigenvalue $S_\al$. Suppose there is a MNE $S$ associated with an eigenfunction $\phi$ at $\v q=\v Q$, we have
\eqa H_\Ga \sim \sum_{\v k,\v k',m,n}\psib_{\v k+\v Q}\psib_{-\v k}f_m(\v k) \phi_m S \phi_n^* f_n^*(\v k')\psi_{-\v k'}\psi_{\v k'+\v Q}.\nonumber\eea
The divergence of $S$ implies an emerging Cooper pairing at collective momentum $\v Q$ with the pairing function
\eqa \phi(\v k)= \sum_m \phi_m f_m(\v k), \label{eq:gapfunc} \eea
with explicit summation over bilinear labels.
By Cooper mechanism, the most favorable collective momentum is $\v Q=0$ for time-reversal-invariant systems. $\phi(\v k)$ forms an irreducible representation of the little group at $\v Q$, and degeneracy exists if it belongs to a multiplet irreducible representation.
	
Similarly, we can rewrite $H_\Ga$ in terms of $V^{\rm DW}$ (for DW = SDW/CDW)  as
%\begin{widetext}
	\eqa H_\Ga &&\sim \sum\psib_{\v R_0}\psi_{\v R_0+\v r_m}V^{\rm DW}_{mn}(\v R) \psib_{\v R_0+\v R+\v r_n}\psi_{\v R_0+\v R} \nn
    &&\sim \sum_{\v k,\v k',m,n}\psib_{\v k+\v Q}\psi_\v k f_m(\v k) \phi_m S \phi_n^* f_n^*(\v k')\psib_{\v k'}\psi_{\v k'+\v Q},\nonumber\eea
%\end{widetext}
where $V^{\rm DW}_{mn}(\v R) = V^{\rm DW}_{(0,\v r_m),(\v R,\v R+\v r_n)}$.
In the last step we assume the matrix $V^{\rm DW}(\v q)$ has a MNE $S$ associated with the eigenfunction $\phi$ at momentum $\v q=\v Q$. The divergence of $S$ here implies an emerging density-wave order in the PH channel. The structure of the order parameter is described again by the function $\phi(\v k)=\sum_m\phi_m f_m(\v k)$, but now for the PH pair. If it is independent (dependent) of $\v k$, it describes site-local (bond-centered) density-wave order. Coexistence of site-local and bond-centered density-wave can also be captured. Notice that $\v Q=0$ is not generally favorable in the PH channel (unless at a vHS), and for $\v Q\neq 0$ there is degeneracy in $\v Q$'s related by point group symmetry.

We now include the other internal degrees of freedom. To each $\v r_m$ we associate a pair of orbitals $aa'$, a pair of spins $\si\si'$, and a pair of sublattices $ss'$ for the two fermions within the fermion bilinear. (The two sublattice labels are not independent since they are related by the displacement $\v r_m$.) We can group the (orbital, spin, sublattice) into a combined label $\mu=(a,\si,s)$. The leading eigenfunction $\phi^{\mu\mu'}(\v k)$ now becomes a matrix, providing additional informations on pairing of orbitals, spins and sublattices in the order parameter (applicable for both PP and PH channels). 

\subsection{Truncation of fermion bilinears}
\begin{figure} %[here]
	\includegraphics[width=\columnwidth]{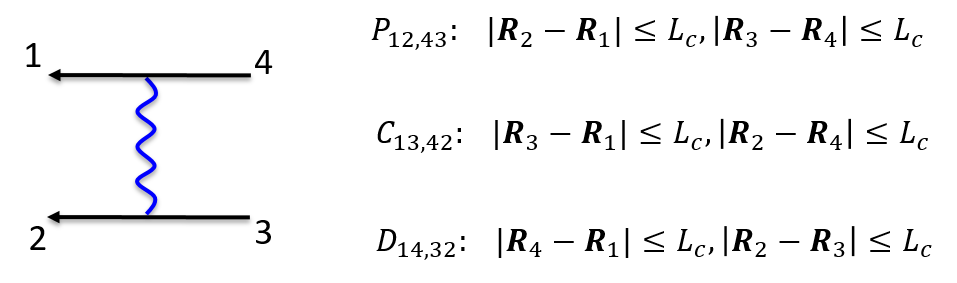}	
	\caption{Illustration of assignment of a 4-point vertex $\Ga_{1234}$ into the three scattering channels according to the truncation length $L_c$. A vertex is an overlap if the assignment can be made in two or all of the three channels.}\label{fig:overlap}
\end{figure}

The flow equation in the form of Eq.\ref{eq:flow} is still not useful if all fermion bilinears, the number of which diverges in the thermodynamic limit, are to be included. We argue that only a finite set of bilinears (in terms of the internal degrees of freedom within the bilinear) are important in a potentially diverging (or singular) scattering mode, the underlying idea of SM-FRG. We observe that if only one out of $P$, $C$ and $D$ is retained in Eq.\ref{eq:flow}, the flow equation reduces to the ladder approximation for $P$, and to the RPA for $C$ and $D$, see below. One would be able to address instabilities in such channels separately. Although subject to serious biases, these approximations do help demonstrate how a generally marginal 4-point vertex could become relevant: by repeated and coherent scattering of fermion bilinears. When an eigen scattering mode becomes singular (or has a diverging eigenvalue), it signals an instability of the normal state, and the associated eigenfunction describes the emmerging order, which is a linear combination of the fermion bilinears. Since in all known examples of ordered state, the order parameter is local or short-ranged, such as local $s$-wave pairing, $d$-wave pairing on bond, site-local density-waves, etc., it is perceivable that the most important fermion bilinears (in the respective scattering channel) that would enter a singular scattering mode are local or short-ranged. (This is not withstanding possible long-range correlations between fermion bilinears.) In fact, unless it is attractive already at the tree level, a scattering channel could become attractive and singular during RG only by its overlap with the other channel, as is clear in the flow equation. If the overlap is strong and very nonlocal, the doner channel must have developed strong nonlocal correlations and hence may diverge even faster. Therefore, we can truncate the relative spatial range in the bilinears entering $P$, $C$ and $D$, say up to a length scale $L_c$. (The setback distance between fermion bilinears is unlimited.) Fig.\ref{fig:overlap} shows how in real space a 4-point vertex is ascribed to $P$, $C$ or $D$. A vertex is overlapped if it can be ascribed to two or all of the truncated scattering channels.
In fact, limiting the fermion bilinears to site-local spin-density and $d$-wave pairing on first-neighbor bonds proved already successful for the Hubbard model describing cuprates.\cite{Husemann_Hubbard_model}

\begin{figure} %[here]
	\includegraphics[width=0.9\columnwidth]{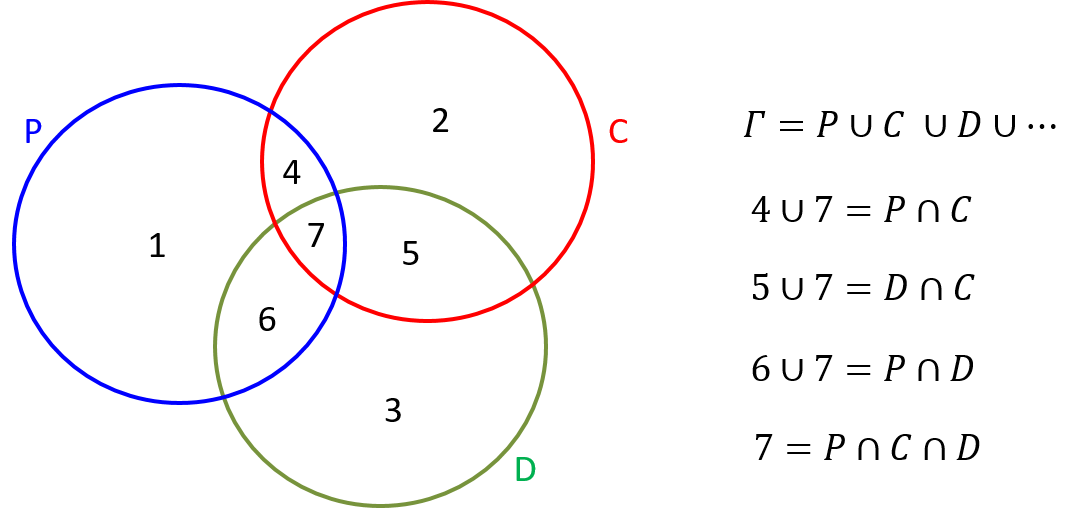}	
	\caption{The relation among truncated $P$, $C$, $D$ and the full vertex $\Ga$ in terms of set theory. Without truncation, $P$, $C$ and $D$ are fully overlapped and are simply aliases of $\Ga$. After truncation the overlaps are partial but sufficient to deal with potentially singular scattering modes in PP and PH channels.}\label{fig:pcd}
\end{figure}

Without truncation, $P$, $C$ and $D$ are simply aliases of $\Ga$. With truncation they capture only parts of $\Ga$ which are most important according to the above arguments. The overlaps between them, shown schematically in Fig.\ref{fig:pcd}, must be retained to treat instabilities in PP and PH channels on equal footing. This is achieved by integrating Eq.\ref{eq:flow} step by step, and after each step, $P$, $C$ and $D$ are reassigned by $\Ga$ according to Eq.\ref{eq:pcd}, or Fig.\ref{fig:overlap}. This scheme is asymptotically exact for parametrization of $\Ga$ if the truncation range is enlarged. A finite truncation makes the calculation feasible, and is sufficient to capture general PP and PH order parameters defined on site and on short-ranged bonds up to $L_c$. Notice that the loop integration is performed in momentum space, but the overlaps between the scattering matrices are handled most conveniently in real space, since overlaps are restricted by the truncation length in fermion bilinears, see Fig.\ref{fig:overlap}.

The full matrix pairing function for the model studied in the main text can be written as $\phi(\v k)\sim i\tau_2$ in the orbital basis. This indicates that the dominant pairing occurs between local orbitals, even though the truncation length $L_c$ is chosen as the length of the second-neighbor bonds, showing $L_c$ is sufficiently large for our purpose. 

\subsection{FRG in the weak coupling limit}
If the channel overlaps between $P$, $C$ and $D$ are ignored, Eq.\ref{eq:flow} reduces to three equations for these scattering matrices, and they can be solved exactly in terms of $V^{\rm SC/SDW/CDW}$. Since the starting interaction is local in real space, we can also limit the fermion bilinears to be local ones. In this basis, we obtain, in matrix form,
\eqa && V^{\rm SC}(\v q) =\frac{V^{\rm SC}_\infty}{1 + V^{\rm SC}_\infty \chi_{pp} (\v q)},\nn
&& V^{\rm SDW}(\v q) =\frac{V^{\rm SDW}_\infty}{1+V^{\rm SDW}_\infty \chi_{ph}(\v q)},\nn
&& V^{\rm CDW}(\v q) =\frac{V^{\rm CDW}_\infty}{1+V^{\rm CDW}_\infty\chi_{ph} (\v q)}.\label{eq:rpa}\eea
Here $\chi_{pp/ph}(\v q)=\int_{\La_0}^\infty d\La \chi'_{pp/ph}(\v q,\La)$ is the susceptibility matrix in the local bilinear basis at the collective momentum $\v q$, contributed by quasiparticles above the energy scale $\La_0$, see Eq.\ref{eq:chi}. We can solve $P$, $C$ and $D$ as
\eqa P=V^{\rm SC}, \ \ C = - V^{\rm SDW}, \ \ D = \frac{1}{2} \left(V^{\rm CDW}- V^{\rm SDW}\right).\nonumber\eea
The effective interaction between Cooper pairs $(\v k,1; -\v k,2)$ and $(\v k',4;-\v k',3)$ can be most conveniently written as, in the bilinear basis,
\eqa V_{12,43}(\v k, \v k') &&= P_{12,43}(0) + D_{14,32}(\v k-\v k') \nn
&&+ C_{13,42}(\v k + \v k')-2[P_\infty]_{12,43}(0). \label{eq:flex}\eea
where the last term subtracts over counting. Notice that starting from the second order in the bare interactions, $P$, $C$ and $D$ collect contributions from independent Feynman diagrams under the given approximation, and this is why they all appear on the rhs of the above equation. To see the connection to the usual expression for the one-band Hubbard model, just substitute $V^{\rm SC}_\infty = V^{\rm CDW}_\infty = - V^{\rm SDW}_\infty=U$ in Eq.\ref{eq:rpa}.  Equation \ref{eq:flex} reflects the fact that fluctuations in the PH channel contribute (attractive or repulsive) effective pair-pair interaction, a mechanism referred to as fluctuation-exchange (FLEX).\cite{FLEX} We remark that even though the bilinears are local in $P$, $C$ and $D$, Eq.\ref{eq:flex} effectively reintroduces long-range bilinears for $V$ from $C$ and $D$ (through the setback displacement between fermion bilinears therein). 

We can now project the above pair interaction onto the band basis to form $V(\v k,\v k')$, and use the Eliashberg equation
to get the leading pairing function, as discussed in the main text. Conceptually this may also be termed a two-step FRG, in the sense that the Eliashberg theory is equivalent to FRG flow in the Cooper channel, using $V(\v k,\v k')$ as the initial pairing interaction at the scale $\La_0$. For repulsive local interactions, $V^{\rm SC/CDW}$ is screened and unimportant in FLEX. However, negative divergence may appear in $V^{\rm SDW}$ too soon versus the strength of the bare interaction (via the Stoner mechanism). 
Consequently, FLEX  works in the weak coupling limit where no divergences appear in RPA.

\bibliography{TBG}

\end{document}